\documentstyle[aps,eqsecnum,epsfig]{revtex}

\newcommand{\bq}{\begin{equation}}
\newcommand{\eq}{\end{equation}}
\newcommand{\bqn}{\begin{eqnarray}}
\newcommand{\eqn}{\end{eqnarray}}
\newcommand{\nb}{\nonumber}
\newcommand{\lb}{\label}

\begin{document}
\twocolumn[\hsize\textwidth\columnwidth\hsize\csname
@twocolumnfalse\endcsname
\title{Critical Phenomena in Gravitational Collapse: The Studies So Far}
\author{Anzhong Wang \thanks{E-mail  address:
wang@dft.if.uerj.br}}
\address{ Departamento de F\' {\i}sica Te\' orica,
Universidade do Estado do Rio de Janeiro, 
Rua S\~ ao Francisco Xavier 524, Maracan\~ a,
20550-013 Rio de Janeiro~--~RJ, Brazil }
 
\date{ \today }

\maketitle

\begin{abstract}

Studies of black hole formation from gravitational collapse  have revealed
interesting non-linear phenomena  at the threshold of black hole formation.
In particular, in 1993 Choptuik studied the collapse of a massless scalar
field with spherical symmetry and found some behaviour, which is quite
similar to the critical phenomena well-known in {\em Statistical Mechanics} and
{\em Quantum Field Theory}.    Universality and echoing of the critical
solution and power-law  scaling of the black hole masses  have given rise to
the name {\em Critical Phenomena in Gravitational Collapse}. Choptuik's results
were soon confirmed   both numerically and semi-analytically, and have extended
to various other matter fields.  

{\mbox{\hspace{.3cm}}} In this paper, we shall  give a brief
introduction to this fascinating and relatively new area, and
provides an updated publication list. An analytical  ``toy"
model of critical collapse is presented, and  some current investigations
are given.

\end{abstract}

\vspace{.4cm}



\vskip2pc]

\section{INTRODUCTION}
 
Gravitational collapse of a realistic body has been one of the most important
and thorny subjects in General Relativity (GR) since the very early times of
GR   \cite{Joshi93}. The collapse generally has four kinds of possible
final states.  The first is simply the halt of  the process in a
self-sustained object, such as,  stars. The second is the dispersion of the
collapsing object and  finally leaves behind  a flat spacetime.
 The third is the formation of black holes with outgoing gravitational and matter
radiation, while  the fourth is the formation of naked
singularities. For the last case, however, the cosmic censorship hypothesis
\cite{Penrose69}  declares that  these naked singularities do not occur in
Nature.   

Due to the mathematical complexity of the Einstein field
equations, we are frequently forced to impose some symmetries on the
concerned system in order to make the problem tractable.  Spacetimes
with spherical symmetry are one of the cases. In particular,
gravitational collapse of a minimally coupled massless scalar field in
such spacetimes was studied both analytically \cite{Chris86} and
numerically \cite{GP87}, and some fundamental theorems were
established.  Quite recently this problem has further attracted
attention, due to Choptuik's discovery of critical phenomena that were
hitherto unknown \cite{Chop93}. As a matter of fact, it is so attractive that
Critical Phenomena in Gravitational Collapse has  already been a very
established sub-area in GR, and several comprehensive review articles already
exist \cite{AEsum,Gun96,Bizon96,Horne96,Chop98,Gun98,Lie00,Gun00}. 

In this paper, a summary of an invited talk given at  {\em the XXI Brazilian
National Meeting on Particles and Fields}, we shall first briefly review the
subject and give an updated list of publication in this area.
This will be done in   Sec. II, while  in Sec. III, we shall present an
analytic ``toy" model of a collapsing massless scalar field. The word ``toy"
model here means that the model doesn't really represent critical collapse,
since the perturbations of the corresponding ``critical" solution have more
than one unstable mode. However, it does have all the main features of
critical collapse. Since so far, no any  critical solution is
known explicitly in a  close  form, this toy model still
serves as a good illustration to critical phenomena in gravitational collapse.
The paper is closed by Sec. IV, in which some  current
investigations in this fascinating area are given.  

\section{Critical Phenomena in Gravitational Collapse}

Starting with spherical spacetimes,
\bq
\lb{1.1}
d{s}^{2} = - {\alpha}^{2}(t, r)dt^{2} 
+ {a}^{2}(t, r)dr^{2} + r^{2}d\Omega^{2},
 \eq
where $d\Omega \equiv d\theta^{2} + \sin^{2}\theta d\varphi^{2}$, and
$\{x^{\mu}\} = \{t, r, \theta, \varphi\}$ are the usual spherical 
coordinates, Choptuik \cite{Chop93} investigated gravitational collapse of a
massless scalar field, $\phi$, which satisfies the  Einstein-scalar
field equations,
 \bqn
\lb{1.2}
R_{\mu\nu} &=& \kappa \phi_{,\mu} \phi_{,\nu},\nb\\
\Box \phi &=& 0,
\eqn
where $R_{\mu\nu}$ denotes the Ricci tensor, $\kappa \equiv [8\pi G/c^{4}]$ is
the gravitational coupling constant, $(\;)_{,\mu} = \partial
(\;)/\partial x^{\mu}, \; \Box \equiv
g^{\alpha\beta}\nabla_{\alpha}\nabla_{\beta}$, and $\nabla_{\alpha}$ denotes
the covariant derivative.  Once an initial smooth configuration of the massless
scalar field is given, these equations uniquely determine the later evolution
of the spacetime and the scalar field \cite{Chris86}. Let the initial
distribution of the massless scalar field be parameterized smoothly by a
parameter $p$ that characterizes the strength of the initial conditions, such
that the collapse of the scalar field with the initial data $p > p^{*}$  forms
a black hole, while the one with  $p < p^{*}$ does not. A simple example  
is the gaussian distribution of the massless scalar field 
\bq 
\lb{1.3} 
\phi(t_{0}, r) = \phi_{0}
\left(\frac{r}{r_{0}}\right)^{3}\exp\left\{ - \left(\frac{r -
r_{0}}{\delta}\right)^{q}\right\}, 
\eq
where  $t_{0}$ denotes the initial time of the collapse, and $\phi_{0}, \;
r_{0}, \; \delta$, and $q$ are constants [See Fig. 1].

\begin{figure}[htbp]
  \begin{center}
    \leavevmode
    \epsfig{file=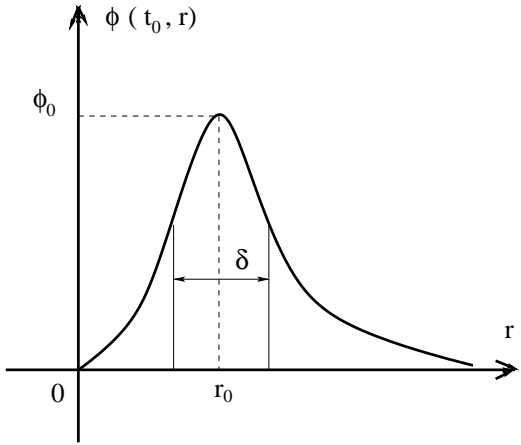,width=0.4\textwidth,angle=0}
\caption{The initial configuration of the massless scalar field at $t = t_{0}$
given by Eq.(\ref{1.3}) in the text. It actually represents a collapsing
spherical shell, made of the massless scalar field, with its thickness
$\delta$ and centralizing at the radius $r = r_{0}$. $\phi_{0}$ represents the
amplitude of the wave packet.}     
 \label{fig1}  
 \end{center} 
\end{figure}

\noindent In this case, Choptuik found that the parameter $p$ can be any of the
four parameters,
\bq
\lb{1.4}
p = \{\phi_{0}, \; r_{0},\; q, \; \delta\},
\eq
that is, fixing any three of the four parameters, for example,  $r_{0},
\; \delta$ and $q$,  and leaving only  $\phi_{0}$ change, we shall obtain a
family of initial data, $S[\phi_{0}]$. For this family of
initial data Choptuik found that there exists a critical value  $\phi^{*}_{0}$
such that when $\phi_{0} > \phi^{*}_{0}$ the collapse {\em always} forms black
holes, and when $\phi_{0} < \phi^{*}_{0}$ the massless scalar field first
collapses, then disperses to spacelike infinity, and finally leaves a flat
spacetime behind without forming any kind of spacetime singularities.
When $\phi_{0} \approx \phi_{0}^{*} + \Delta \phi_{0}$, where $\Delta 
\phi_{0}$ is very small,  after certain time all the collapses are quite
similar and approach to the critical one ($\phi_{0} = \phi_{0}^{*}$). But,
at the very end, the collapse will suddenly  runs
away from the critical one, by either forming black holes or dispersing to
infinity, depending on the signs of $\Delta \phi_{0} $.  Choptuik found
that  for the configuration of Eq.(\ref{1.3})   there are four different
families of initial data, $S[\phi_{0}],\; S[r_{0}],\; S[\delta]$ and $S[q]$,
which all exhibit the above behaviour.

In addition to these four, Choptuik also studied many
others and found that for all the families that behave  as above,
the so-called {\em generic smooth families} of initial data,  the  critical
solutions are identical, or in another word, {\em universal}. Moreover, the
critical solution is also    {\em periodic}, that is,
\bq
\lb{1.4a}
{A}^{*}(\xi, \tau) = {A}^{*}(\xi, \tau + \triangle),
\eq
where ${A}^{*} = \{{\alpha}^{*}, {a}^{*}, 
{\phi}^{*}\}$, and 
\bq
\lb{1.5}
\tau = \ln\left(\frac{t}{r_{0}}\right),\;\;\;
\xi = \ln\left(\frac{r}{t}\right) - \xi_{0}(\tau), 
\eq
with $r_{0}$ being a dimensionful constant, and $\xi_{0}(\tau)$ a  periodic
otherwise arbitrary function with period $\triangle$.  The  constant
$\triangle$ is a {\em dimensionless } constant, which   was numerically
determined as $\triangle \approx 3.447$.  
   
Yet, near the critical solution but with $p > p^{*}$,
the  mass of black holes takes the scaling form
\bq
\lb{1.6}
M_{BH} = K (p - p^{*})^{\gamma},
\eq
where $K$ is a family-dependent constant, but $\gamma$ is another 
{\em dimensionless universal} constant, which   was numerically
determined as  $\gamma \approx 0.37$. 

Universality and echoing of the critical
solution and power-law  scaling of the black hole masses  have given rise to
the name {\em Critical Phenomena in Gravitational Collapse}.

Choptuik's  results were soon confirmed by several independent studies both
numerical \cite{Gun94} and semi-analytical \cite{Gun95}, and have been extended
to other matter fields, such as,

\begin{itemize}

\item  Axisymmetric gravitational waves \cite{AE93}; 

\item  Perfect fluids with the equation of state $p = k \rho$,
where $p$ denotes the pressure of the fluid and $\rho$ the energy
density \cite{EC94,KHA95} and $k$ is a constant;

\item  Quantum black hole formation in  2-dimensional spacetimes \cite{ST94};

\item  Non-linear   $\sigma$-models in two dimensional target
space \cite{HE95};

\item Massless scalar field in Brans-Dicke theory \cite{LC96};
  
\item $SU(2)$  Yang-Mills field \cite{CCB96};

\item  Einstein-Maxwell-scalar fields   \cite{GM96};

\item  Massive scalar field \cite{BCG97};

\item Gravitationally collapsing primordial density fluctuations in the
radiation dominated phase of the early Universe \cite{NJ97};

 \item  $SU(2)$ Skyrme field \cite{BC98};  

\item The collapse of collisionless matter of the Einstein-Vlasov
equations \cite{RRS98,Olab00};  

\item  Topological  domain walls interacting with black holes \cite{Frolov99};

\item The gravitational collapse of massless scalar field in higher
dimensional spacetimes \cite{GCD99};

 \item Non-linear   $\sigma$-models in
three dimensional    target space \cite{Liebling99}; 

\item Gravitational collapse in Tensor-Multi-Scalar and Non-linear Gravity
Theories \cite{Wang99}; 

\item Boson stars \cite{HC00}; 

\item Massless scalar
field coupled with the cosmological constant in (2+1)-dimensional
spacetimes  \cite{PC00}. 

\end{itemize}

In review of all these studies, now the following is clear: 

\begin{description}

\item {(a)} In general the
critical solution and the  two dimensionless constants $\triangle$ and
$\gamma$   are universal only with respect to the same matter field, and 
usually are matter-dependent.  For example, for the collapse of  the
$SU(2)$ Yang-Mills field, it was found \cite{CCB96} that $\triangle \approx
0.74$ and $\gamma \approx 0.2$, while in the case of massless scalar field,
Choptuik found that $\triangle \approx 3.447$ and $\gamma \approx 0.37$.

\item{(b)} The critical
solutions can have discrete self-similarity (DSS) \cite{DSS}  or continuous
self-similarity (CSS) \cite{CSS}, or none of them, depending on the
matter fields and regions of  the initial data space.   So far, in all the
cases where the critical solution either has DSS or CSS, black holes form {\em
always} starting   with zero mass, and   
take the form of Eq.(\ref{1.6}), the so-called Type II collapse, while in the
cases in which the critical solution has neither DSS nor CSS, the formation
always turns on with a mass gap, the so-called Type I collapse, 
corresponding, respectively,   to the second- and first-order phase
transitions in Statistical Mechanics \cite{Golden92}. 

\item{(c)}  The universality of the critical solution and the exponent
$\gamma$ now are well understood in terms of perturbations of
critical solutions \cite{KHA95},   while the one of   $\triangle$ still remains
somewhat of a mystery. The former is  closely related to the fact that the
perturbations of the critical solution has only one unstable  mode. This
property now is considered as the main criterion for a solution to  be critical
\cite{Gun00}. 

\end{description}

To understand the last property better, let us consider the phase space, 
that is, consider GR as an infinite-dimensional dynamic system. If we make  
a ($3+1$) split of the spacetime, for example, following the Arnowitt, Deser,
and Misner  (ADM) decomposition, we will find that the dynamic quantities will
be the induced spatial  three metric, the extrinsic curvature, and the
matter distribution.   Then, the phase space will consist of {\em all the
possible} three metrics, extrinsic curvature, and  
configurations of the matter fields. For the case of massless scalar field, 
from the no-hair theorem of black holes \cite{Bekenstein98}, we know that the only
stable black hole solution of the Einstein-scalar field equations is the
Schwarzschild black hole with a constant massless scalar field. Except for
this black hole, another stable state is the Minkowskian spacetime.  
Of course, we also know that the collapse of a massless scalar field can form
naked singularities, too, but so far we don't know if they are stable or not
\cite{Joshi93}. At this point, we shall adopt the point of view of  the cosmic
censorship conjecture \cite{Penrose69}, and assume that they are not stable.
Otherwise, there may exist two more critical solutions that separate,
respectively,  black holes from naked singularities, and flat spacetimes from
naked singularities. However, this doesn't affect our following discussions if
we are restricted only to the boundary between black holes and flat spacetimes,
and the analysis can be easily extended to other boundaries.

\vspace{1.cm}

\begin{figure}[htbp]  
 \begin{center}      
\leavevmode    
\epsfig{file=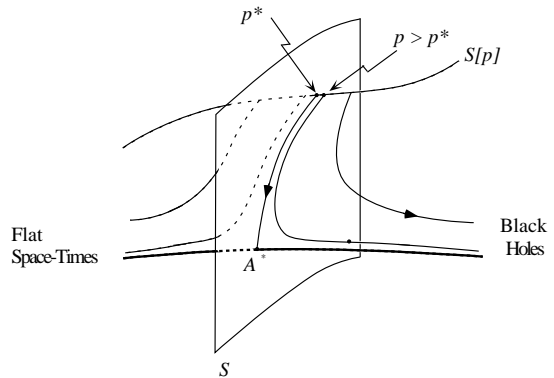,width=0.4\textwidth,angle=0} 
\caption{The phase space of the dynamic system of the Einstein-scalar field
equations. The hypersurface $S$ is the critical surface of codimension one,
which separates the basin of black holes from the basin of flat
spacetimes. A generic smooth family of initial data $S[p]$ always passes
the two basins  at the critical point $p = p^{*}$ on the critical hypersurface.
All the initial data on the hypersurface will collapse to the critical solution
$A^{*}$ that is a fixed point on the hypersurface when it has CSS and a fixed
cycle when it has DSS. All details of initial data are soon washed out
during the collapsing process, and the collapse with initial data near the
critical point will be very similar to the critical collapse. This
similarity can be last almost to the fixed point $A^{*}$, whereby the one
unstable mode suddenly draws the collapse either to form black hole or a
flat spacetime, depending on whether $p > p^{*}$ or $p < p^{*}$. }     
\label{fig2}     
\end{center} 
\end{figure}

\noindent Then, we can see that the phase space can be divided into two
attractive basins. Inside the  dispersion basin, the flat spacetimes 
  with different constant values of the scalar field, are attractive
fixed points, while inside the basin of black holes,   the Schwarzschild 
black holes with different masses   are the attracting fixed points.  The
boundary between the two attractive basins is called the critical surface, and
the critical solution always lies on it. Since it has only one unstable
mode, this surface must be a hypersurface of codimension one, that is, one
dimension less than the original infinite-dimensional phase space. By
definition, a phase space trajectory never leaves this hypersurface, if it is
initially    on  it, but approaches to the critical solution, which is a
fixed point on this hypersurface if the critical solution is continuous
self-similar, or a  fixed cycle if  the critical solution is discrete
self-similar \cite{Gun00}. Within the complete phase space, the critical
solution is an attractor of codimension one, i.e., it has an infinite number
of decaying perturbation modes tangential to the critical hypersurface and a
single growing mode perpendicular to the  hypersurface. Any trajectory
beginning near the critical hypersurface, but not necessarily near the
critical point (or cycle in the DSS case), first moves parallel to the
hypersurface and goes down  almost   to the critical point (or
cycle), then is suddenly drawn away  by the single unstable mode in the
perpendicular direction,  and finally ends up at one of the  fixed points, by
either forming a black hole or a flat spacetime.  During the
dynamic process, all details of the initial data are quickly washed away,
except for the distance from the black hole threshold. Therefore, for the both
super-critical ($p > p^{*}$) and sub-critical ($p < p^{*}$) collapse, there
exists a domain, $ p^{*} - \triangle p \le p \le p^{*} + \triangle p$, in the
phase space,  in which the collapse is very similar to the critical one during
certain period of times  [See Fig.2].

 \section{Critical Collapse of Massless Scalar Field: An Analytic Toy Model}

 In this section, we shall present a class of analytic solutions of the
Einstein-scalar field equations, which represents gravitational collapse of a
wave packet consisting of massless scalar field \cite{WO97}. This class of
solutions was first discovered by Roberts  \cite{Roberts89} and later studied
by several authors in the context of critical collapse \cite{Brady94}. As we
shall show below, these solutions possess most of the features of critical
phenomena, although they don't exactly represent  critical collapse, because
the solution that separates the formation of black holes from that of flat
spacetimes has more than one unstable mode \cite{Frolov97}.  It is exactly in
this sense, we refer these solutions as representing a    ``toy"
model of critical collapse.

The Roberts solutions are given by \cite{Roberts89}
\bq
\lb{eq1}
ds^{2} = -  G(u,v)du dv + r^{2}(u, v) d^{2}\Omega, 
\eq
where $u$ and $v$ represent two null coordinates, in terms of which the
metric coefficients   and the corresponding massless scalar field $\phi$ are 
given, respectively, by  
\bqn
\lb{eq2}
r(u, v) &=& \frac{1}{2}\left( u^{2}  - 2uv + 4 b_{2}v^{2}
\right)^{1/2},\nb\\
G(u, v ) &=& 1,\\
\lb{eq3}
\phi(u, v ) &=& \pm \frac{1}{\sqrt{2}}\ln\left|\frac{(u - v) - (1 -
4b_{2})^{1/2}v} {(u - v) + (1 - 4b_{2})^{1/2}v}\right|,
\eqn
where $b_{2}$ is an arbitrary constant. Note that the notations used here
closely follow the ones used in \cite{WO97} but slightly  different from the
ones used in \cite{Roberts89}. 
From Eq.(\ref{eq2}) it can be easily shown that the local mass function
\cite{PI90} is given by  
\bq
\lb{eq5}
m(u, v) \equiv \frac{r}{2}\left(1 - r_{,\alpha}r_{,\beta} 
g^{\alpha \beta}\right) = - \frac{(1 - 4b_{2})uv}{8r}, 
\eq
which is zero on the hypersurface $v = 0$ and negative for $ u, \; v <
0$. Thus, to have a physically reasonable spacetime we need to
restrict the above solutions valid only in the region $u \le 0,\; v \ge 0$.
Since the mass is zero on the hypersurface $v = 0$, we may   join the above
solutions across the hypersurface $v = 0$ with a Minkowskian spacetime. As
shown in \cite{WO97}, this is possible if the metric in the region $v
\le 0$ takes the form of Eq.(\ref{eq1}) but with the metric coefficients
and the massless scalar field being given by
\bqn
\lb{eq2a}
r(u, v) &=& a(v) - \frac{1}{2}u - a(0),\nb\\
G(u, r) &=&   2 a'(v),\;\;\;\; \phi = 0, \; (v < 0),
\eqn
where $a(v)$ is an arbitrary function subject to $a'(v) > 0$ and $a'(0) =
1/2$, and a prime denotes the ordinary differentiation with respect to the
indicated argument. For such a matching, it can be shown that  the
hypersurface $v = 0$ is free of any kind of matter and represents a
boundary surface \cite{Israel66}. The region $v < 0$
is Minkowskian  [See Fig.3].

\begin{figure}[htbp]  
 \begin{center}      
\leavevmode    
\epsfig{file=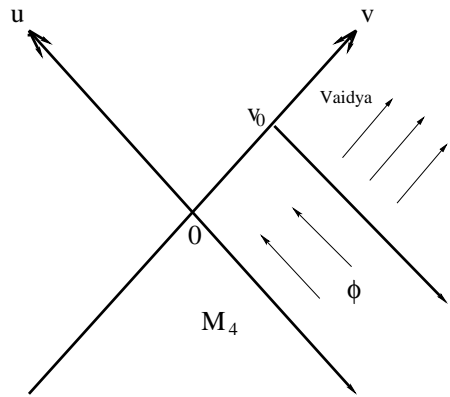,width=0.4\textwidth,angle=0} 
\caption{The spacetime in the ($u,\; v$)-plane. The
region  $v < 0$ is Minkowskian, while the region
 $u \le 0,\; v \ge 0$ represents gravitational collapse of the
massless scalar field. (a) When $b_{2} < 0$, the scalar wave collapses into a
spacetime singularity at $ u = - [(1 - 4b_{2})^{1/2} - 1]v <
0$, which is preceded by  an apparent horizon located at $u = -  4|b_{2}|v <
0$.  (b) When $b_{2} = 0$, the spacetime singularity coincides with the
apparent horizon  on $u = 0$ which is null. (c) When  $0 < b_{2} < 1/4$,   the
massless scalar field first collapses and then disperses into infinity,
and finally  leaves  a Minkowskian spacetime
behind in  the region $u, \; v > 0$. }      
\label{fig3}     
 \end{center}  
\end{figure}  

On the other hand,   from Eqs.(\ref{eq2}) and (\ref{eq3}) it can be also shown
that the spacetime in the region $u < 0, v > 0$  represents a collapsing
massless scalar wave. When $b_{2} < 0$, the scalar wave collapses into a
spacetime singularity on the hypersurface $ u = - [(1 - 4b_{2})^{1/2} - 1]v$,
which is preceded by  an apparent horizon at $u = 4b_{2}v$.  Thus, the
corresponding solutions represent the formation of black holes. When $b_{2}
= 0$, the singularity coincides with the apparent horizon  on the null
hypersurface $u = 0$. When  $0 < b_{2} < 1/4$, it can be
shown  that  the massless scalar field first collapses and then disperses into
infinity, without forming black holes, but instead,  leaves  a Minkowskian
spacetime behind, which now is represented by the region, $u, v > 0$, in which
the metric takes the form of Eq.(\ref{eq1}), but with
\bqn
\lb{eq4}
G(u, v) &=& 4b_{2}^{1/2} b'(u) , \;\;\;\; r = b_{2}^{1/2}\;v 
- b(u) + b(0), \nb\\
\phi(u, v) &=& \pm \frac{1}{\sqrt{2}}\ln
\left[\frac{1 + (1 - 4b_{2})^{1/2}}{1 - (1 - 4b_{2})^{1/2}}\right], \nb\\
& & \;\;\;\;\;\;\;\; (0 < b_{2} < 1/4, \; u, v > 0), 
\eqn
where $b(u)$ is an arbitrary function, subject to $b'(u) > 0, b'(0) =
1/(4b_{2}^{1/2})$. One can show that the hypersurface $u = 0, v >
0$ is also free of any kind of matter and represents a
boundary surface.

In the case $b_{2} < 0$, where black holes are formed, Eq.(\ref{eq5}) shows
that   on the apparent horizon $u = - 4|b_{2}|v$ the mass becomes unbounded
as $v \rightarrow + \infty$.  In order to have black holes with finite mass,
we shall follow \cite{WO97} first to cut the spacetime along the hypersurface
$v = v_{0}$ and then join the region $ 0 \le v \le v_{0}$ with an
asymptotically flat region. To model the out-going radiation of the
massless scalar field, we shall choose the  region $v \ge v_{0}$ as
described by the Vaidya solution \cite{Vaidya51},   
\bqn
\lb{eq12}
ds^{2} &=& - \left(1 - \frac{2 m(U)}{r}\right) dU^{2}  + 2dUdr  + r^{2}
d^{2}\Omega, \nb\\
& & \;\;\;\;\;\; \; (v \ge v_{0}),
\eqn
where $U$ is the Eddington retarded time, which is in general the
function of $u$ appearing in Eq.(\ref{eq1}), and $m(U)$ is the local mass  of
the out-going Vaidya dust.  The corresponding energy-momentum tensor is
given by
\bq
\lb{eq14}
T^{+}_{\mu\nu} = - \frac{2}{r^{2}}\frac{dm(U)}{dU} \delta^{U}_{\mu}
\delta^{U}_{\nu},\; (v \ge v_{0}).
\eq
The hypersurface $v = v_{0}$ in the coordinates $\{x^{\mu}\} = \{U, \; r, \;
\theta, \; \varphi\}$ is given by 
\bq
\lb{eq14a}
\frac{dU(r)}{dr} = - \frac{2r}{r - 2m(U)}, \; (v \ge v_{0}).
\eq
Then, it can be shown that the junction conditions
on the hypersurface $v = v_{0}$ require
\bqn
\lb{eq28}
 M(r)  &\equiv& \left. m(U)\right|_{v = v_{0}} =  \frac{1}{r}\left[
p (4{p^{*}}^{2} + r^{2})^{1/2} -  2 {p^{*}}^{2}\right],  \nb\\
v_{0} &=& \frac{4{p^{*}}^{2}}{p},
\eqn
where $p$ is the integration constant, and
\bq
\lb{eq29}
p^{*} \equiv \frac{(1 - 4b_{2})^{1/2}}{4}v_{0}.
\eq
For the details, we refer readers to \cite{WO97}. Since for the above
matching, the hypersurface $v = v_{0}$ is free of matter, the function $M(r)$
represents the total mass of the collapsing wave packet filled in the
region $0 \le v \le v_{0}$.  At the
past null infinity, Eq.(\ref{eq28}) shows that 
\bq
\lb{eq30} M(r \rightarrow + \infty) = p, 
\eq
that is, the parameter $p$ in the present case represents the total initial
mass of the massless scalar wave packet  with which it starts to
collapse. 

As $r \rightarrow 0^{+}$, from Eq.(\ref{eq28}) we can see that  $M(r)$  
behaves as
\bq
\lb{eq31}
M(r) \rightarrow 
\cases{ + \infty, & $ p > p^{*}$, \cr
0, &  $ p = p^{*}$, \cr
- \infty, &  $ p < p^{*}$.\cr}
\eq

\begin{figure}[htbp]  
\begin{center}      
\leavevmode    
\epsfig{file=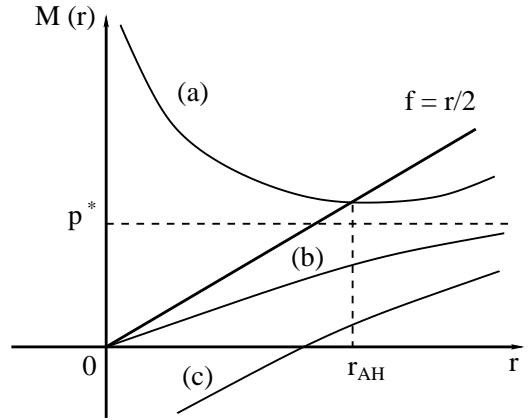,width=0.4\textwidth,angle=0} 
\caption{The mass   $M(r)$ of collapsing spherical shell. The line (a)
corresponds to the case where $p > p^{*}$, in which a black hole is formed,
and its mass is given by $M_{BH} = M(r_{AH})$. The line (b)
corresponds to the case where $p = p^{*}$, while the line (c)
corresponds to the case where $p < p^{*}$.   }       
\label{fig4}     
\end{center}  
\end{figure}

On the other hand, it is well-known that the apparent horizon at $r =
2M(r)$ of the out-going Vaidya solution always coincides with its
future event horizon. Thus, by comparing the mass $M(r)$ with
$r/2$ we can tell whether the collapse forms a black hole or
not,
\bqn
\lb{eq32}
 M(r)  - \frac{r}{2} &=&  \left\{\frac{4p^{*^{2}} + r^{2}}
 {4r^{2}[p + (4p^{*^{2}} + r^{2})]}\right\}^{1/2}\nb\\
& & \;\; \times \left[4(p^{2} - p^{*^{2}}) - r^{2}\right].
\eqn

\noindent Clearly, only when $p > p^{*}$, the scalar field and the null shell
will collapse inside the event horizon at
\bq
\lb{eq33}
r_{AH} = 2 (p^{2} - p^{*^{2}})^{1/2}.   
\eq

 \begin{figure}[htbp]  
 \begin{center}      
\leavevmode    
\epsfig{file=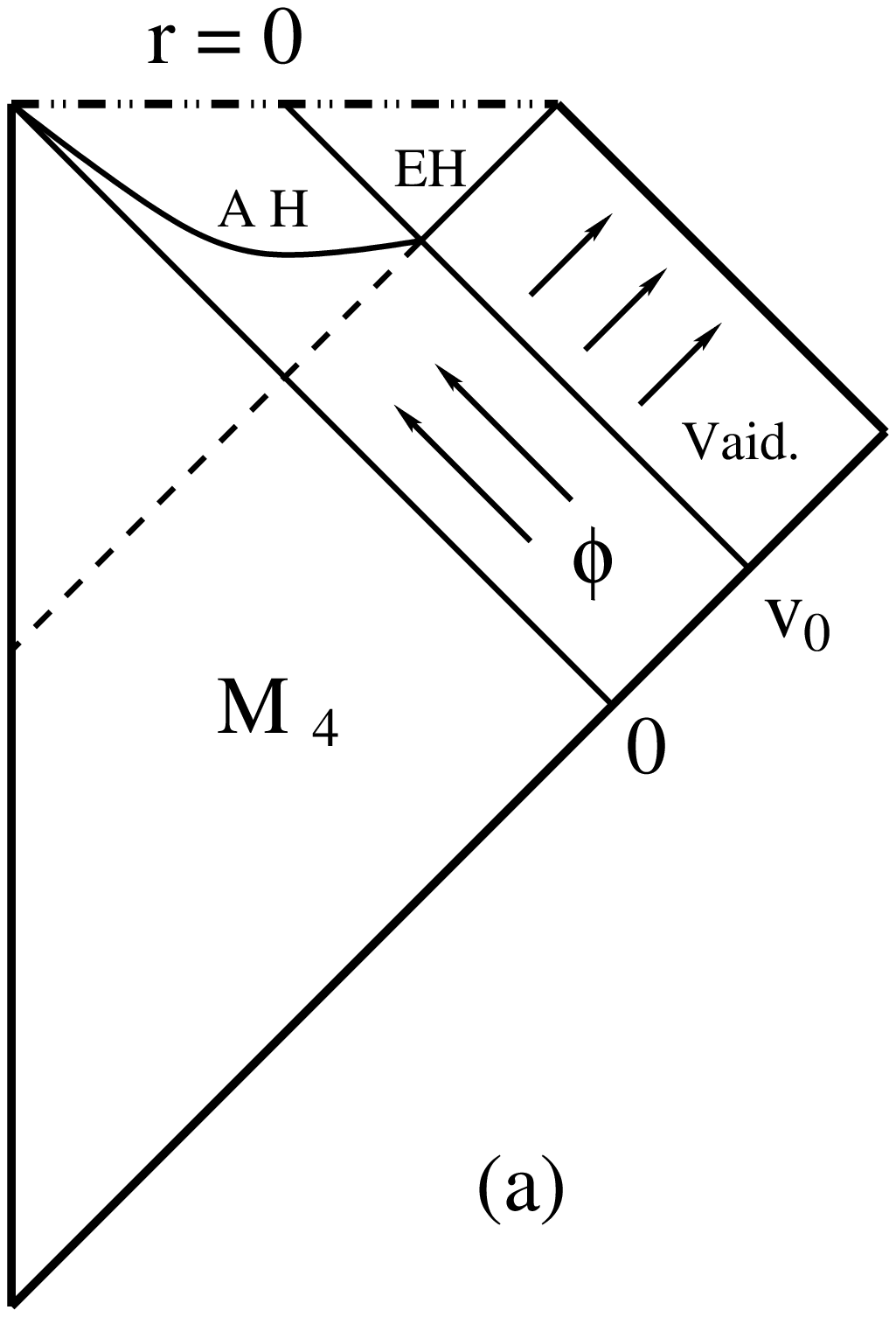,width=0.4\textwidth,angle=0} 
   
\vspace{1.cm}

\epsfig{file=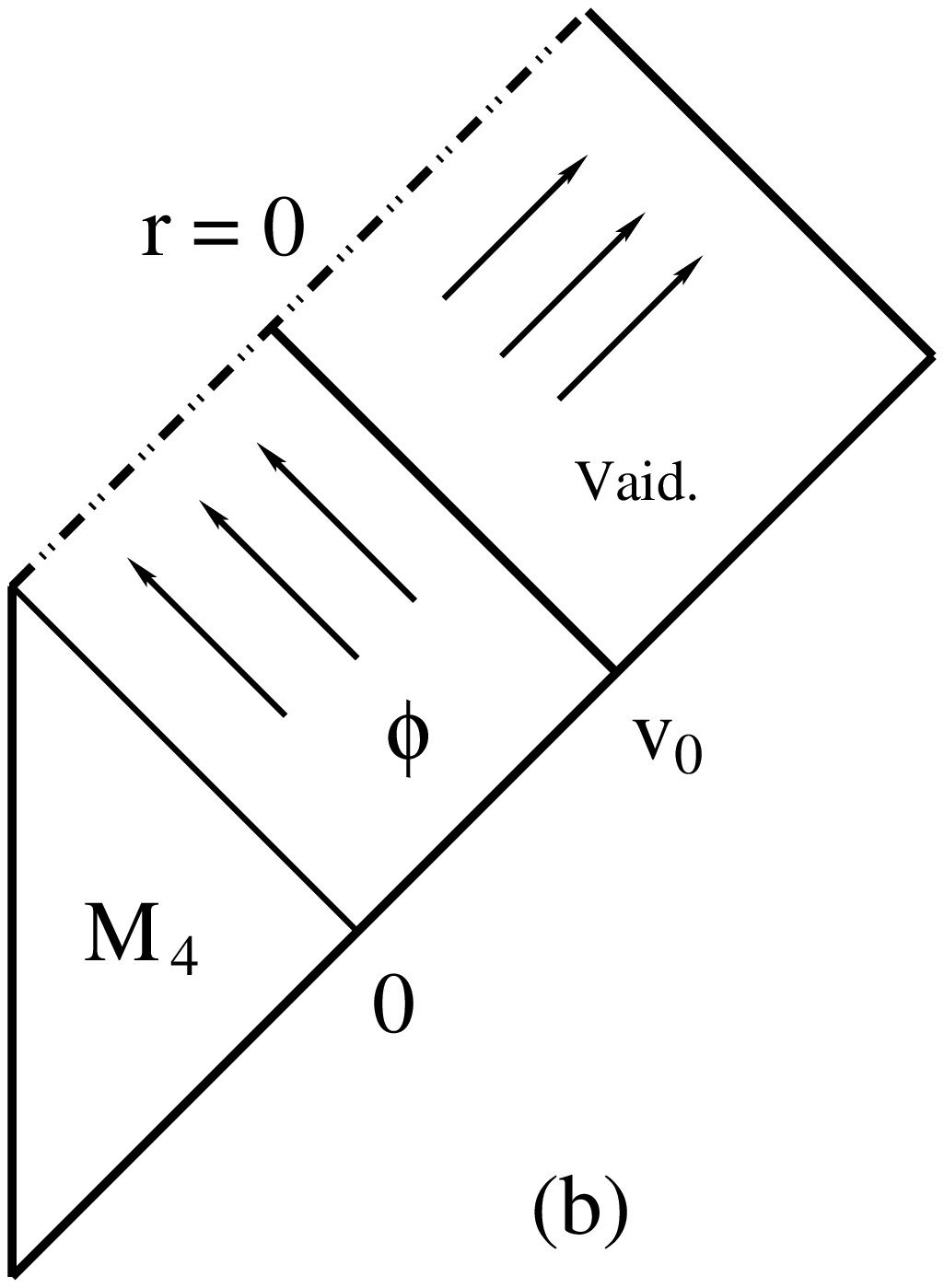,width=0.4\textwidth,angle=0} 
   
\epsfig{file=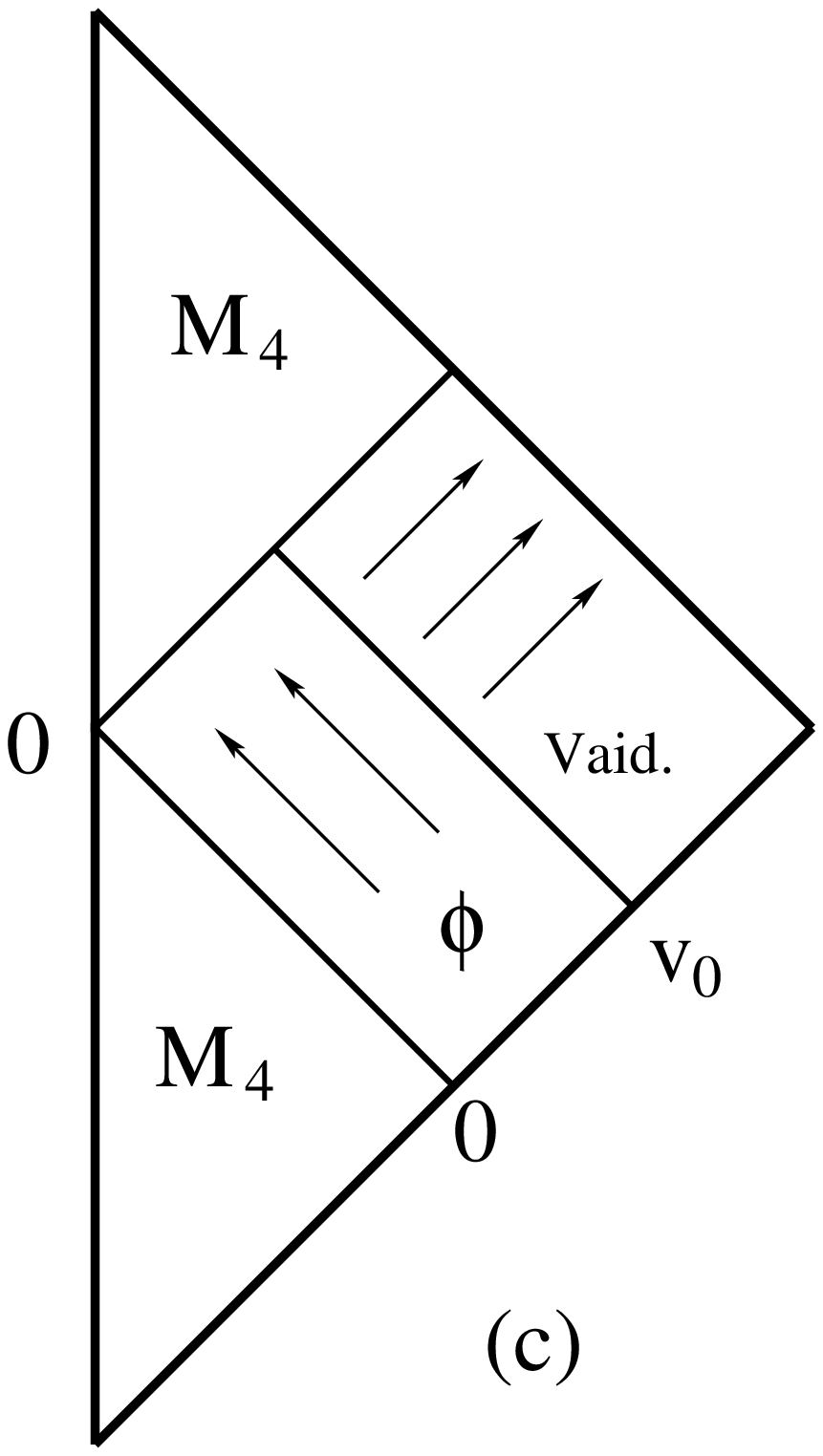,width=0.4\textwidth,angle=0} 
\caption{The corresponding Penrose diagrams. (a) The case where $p >
p^{*}$, in which a black hole is formed, and its mass is given by $M_{BH} =
M(r_{AH})$.  (b) It corresponds to the case where $p = p^{*}$, in which
the spacetime singularity becomes null and coincides with the apparent
horizon.    (c) It corresponds to the case where $p < p^{*}$, in which no
spacetime singularity is formed, instead, when the wave packet of the massless
scalar field collapses to the hypersurface $u = 0$, all of its mass has been radiated
away and nothing is left to collapse, so  the spacetime in the region $u, \; v
\ge 0$ becomes Minkowskian. }       
\label{fig5}     
 \end{center}  
\end{figure}

\noindent When $p = p^{*}$, $M(r) = r/2$ is possible only at the origin,
$r = 0$, where a zero-mass singularity is formed. Thus, the solution
with $p = p^{*}$ represents the ``critical" solution that separates the
supercritical solutions ($p > p^{*}$) from the subcritical ones ($p <
p^{*}$). In the subcritical case, $M(r)$ is always less than
$r/2$, and the collapse never forms a black hole [cf. Fig.4].

In the subcritical case, the region $u, v > 0$ should be replaced by the
Minkowskian solution (\ref{eq4}).  As shown above, the
matching across the hypersurface $u = 0, \; 0 \le v \le v_{0}$ is
smooth, i.e., no matter appears on it. To show that it is also the case
on the hypersurface $u = 0, \; v \ge v_{0}$, which separates the Vaidya
solution (\ref{eq12}) from the Minkowskian one (\ref{eq4}), we first make the
coordinate transformation $U = U(u)$, and then write the metric
(\ref{eq12}) in terms of $u$. Using the results obtained in
\cite{BI91}, one can show that to have a smooth matching we have to
impose the condition
\bqn
\lb{eq33a}
U'(0) = \frac{1}{(4b_{2})^{1/2}},\\ 
\lb{eq33b}
M(r)\left|_{u = 0} = 0\right..
\eqn
Clearly, by properly choosing the function-dependence of $U(u)$,
the first condition (\ref{eq33a}) can be always satisfied. On the other hand,
from Eqs.(\ref{eq2}) and (\ref{eq28}) one can show that the last condition
is also satisfied identically. Therefore, the matching of the Vaidya
solution to the Minkowskian one across the hypersurface $u = 0, \; v \ge
v_{0}$ is always possible for $p < p^{*}$. The corresponding Penrose
diagram for each of the three cases are shown in Fig. 5.

On the apparent horizon $r = r_{AH}$, the total mass of the scalar wave
packet is given by
\bq
\lb{eq34}
M_{BH} = \frac{1}{2}r_{AH} = K (p - p^{*})^{1/2},  
\eq
where $K \equiv (p + p^{*})^{1/2}$. The above expression shows that
the black hole mass takes a power-law form with its exponent $\gamma = 0.5$,
which is different from the DSS case, where Choptuik found $\gamma \approx
0.37$. As we mentioned previously, the above solutions don't really represent
critical collapse, because the ``critical" solution given above has more than
one unstable modes \cite{Frolov97}. Therefore, the different value of $\gamma$
obtained here does not means any contridiction to Choptuik's numerical results.

\section{Current Investigations}

As Critical Phenomena in gravitational collapse is a rather new area in GR,
there are many open problems. In the following we shall mention some of
them.

\subsection{The Effects of Angular Momentum in Critical Collapse}

As we know, the angular momentum plays a very significant role in   black
hole  physics,  and all the realistic bodies, such as, neutron stars, have
non-zero angular momentum. Thus, it is very important to study the effects of
angular momentum on critical collapse. So far, all the studies of critical
phenomena in gravitational collapse have been restricted to spherical case,
except for the works of Abrahams and Evans \cite{AE93} and Alcubierre et al.
\cite{Alcu00}. In \cite{AE93} the authors  studied the collapse of  
axisymmetric purely gravitational waves, and found the type II critical
collapse. However, in this study  the total angular momentum is still
zero.  In \cite{Alcu00} the collapse of  pure Brill type gravitational
waves   in 3D Numerical Relativity was studied and the
critical amplitude for black hole formation was determined. However, due
to the complexity of the problem, no sufficient evidence for critical collapse
was observed.    

In addition to the above, Gundlach and his co-workers \cite{Gunons} studied the
problem using non-spherical perturbations, and in particular found that all the
modes of non-spherical perturbations of the massless scalar field studied
initially by Chpotuik are stable, and, as a result, Chpotuik's critical
solution found in the spherical case may remain critical even in
non-spherical case. He also found that small angular momentum also takes a
scaling form near the critical point but with a different exponent. Besides,
Rein, Rendall and Schaeffer \cite{RRS98} studied the spherical collapse of
collisionless matter that consists of counter-rotating particles, and found
that only Type I critical collapse. This result was further confirmed
by Olabarrieta quite recently \cite{Olab00}. 

Moving from spherically symmetric case to axisymmetric one, the problem
becomes much mathematically involved, and very sophisticated (numerical) 
methods are needed. Choptuik, Hirschmann and Liebling, among others, have been
working on this problem recently \cite{Wang01}, and are expected to report
their results soon.

\subsection {The Quantum Effects on Critical Collapse}

Critical phenomena are actually phenomena in the strong gravitational field
regime, and Quantum effects should be very important for the formation of
black holes with very small mass.  Chiba and Siino \cite{CS97} studied this
problem and showed that the Quantum effects may destroy the type II critical
phenomena, while Ayal and Piran \cite{AP97} showed that they don't, but rather
 shift the critical value $p^{*}$. However, since in both of the two cases the
Quantum effective energy-momentum tensor (EMT)  was taken from two-dimensional
toy model, the consistence of such an  EMT with the four-dimensional
gravitational collapse is still an open question. Recently, Brady and Ottewill
\cite{BO98} calculated the  effective EMT of a conformally coupled scalar
field on the fixed background of the critical solutions of the perfect fluid
with the equation of state $p = k \rho$ in four-dimensional spherical
spacetimes,  and found that when $k < 0.53$,  the  Quantum effects destroy
the type II critical phenomena,  while when $k > 0.53$ their calculations break
down, and a definitive conclusion is still absent.

\subsection {The Application of Renormalization Group Theory to Critical
Collapse}

The Renormalization Group Theory has achieved great success in the
studies of critical phenomena in Statistical Mechanics \cite{Golden92},
and several authors, including Argyres \cite{Arg94}, and Koike, Hara and
Adachi \cite{KHA95}, have pointed out that the time evolution near the
critical solution in gravitational collapse may also be considered as a
renormalization group flow on the phase space of initial data. As a
matter of fact, the analysis of the phase space given in the
Introduction exactly followed this idea. However, this analysis is valid only
in self-similar spacetimes. In order to obtain a full renormalization group,
one needs to generalize them to arbitrary spacetimes, which is turned out not
trivial, as in GR the choice of coordinate systems is completely arbitrary.
Garfinkle and Gundlach \cite{GG98} have   taken some initial steps to this
direction, but a successful application of the Renormalization Group Theory to
critical collapse  still remains as an open question.

Besides the above mentioned problems  many others   are
also under the current investigations, such as, finding more matter fields
that exhibit critical collapse, including universal classes;
applying the analysis of perturbations of black holes to critical
solutions; understanding its physical origin of the constant $\triangle$
\cite{PP96}; finding some possible astrophysical observations of critical
phenomena, and so on.  In particular, it was known for a long time that the
collapse of neutron stars exhibits the type I critical phenomena
\cite{HTWW65}. An important question is that: Does this type I critical
collapse have any observational consequence?  

For further references of critical phenomena in gravitational collapse, we
would like to refer the readers to the review articles
\cite{AEsum,Gun96,Bizon96,Horne96,Chop98,Gun98,Lie00,Gun00}.  

\section*{ACKNOWLEDGMENTS}

The author would like to express his gratitude to M.W. Choptuik, S.L.
Liebling,  J. Pullin, and W.M. Suen for valuable  discussions and suggestions
in critical collapse.  He would also like very
much to thank his collaborators in this area, C.F.C. Brandt,  R.-G. Cai,
E.W. Hirschmann, L.-M. Lin, H.P. de Oliveira, J.F. Villas da Rocha, N.O.
Santos, and Y.M. Wu. The financial  assistance from CNPq is gratefully
acknowledged.

\end{document}